\documentstyle[11pt,newpasp,twoside,epsf]{article}
\def\edcomment#1{\iffalse\marginpar{\raggedright\sl#1\/}\else\relax\fi}
\reversemarginpar

\begin{document}
\title{Structures in Dusty Disks}
\author{J.-C. Augereau}
\affil{Sterrewacht Leiden, P.O. Box 9513, 2300 RA Leiden, The
Netherlands}

\begin{abstract}
Optically thin dusty disks around Main Sequence stars consist of
debris from catastrophic collisions or from low erosion of long-lived
planetesimals. Resolved observations of dusty disks have
systematically evidenced asymmetries and annular structures.  It has
been proposed that some of these structures could be signatures of
undetected exoplanets. In this paper, I review and discuss currently
proposed models to account for the observed structures. These include
the impact of distant exoplanets and/or of stellar companions. The
Solar System serves as a reference case for these approaches and
similarities are pointed out along the paper as well as some
limitations of current modeling efforts.
\end{abstract}

\section{Extra-solar Dusty Disks}
A major legacy of the IRAS and ISO missions is the identification of
mid- and far-infrared excesses for a large fraction of nearby Main
Sequence (MS) stars. The flux calibrator star Vega turned out to be
representative of a class of MS stars surrounded by optically thin
disks of cool solid material. It is now widely accepted that the Vega
phenomenon likely represents a common stage in circumstellar disk
evolution (Lagrange, Backman \& Artymowicz\,2000). This stage follows
the dissipation of the initial massive gas disk which can support
planetary formation. Early after the first IRAS results, Harper et
al.\,(1984) concluded that because the observed grains are short-lived
in the Vega disk, they are either collisional debris or ejected
particles from evaporating exo-comets, implying in both cases a
reservoir of large bodies generally referred to as
planetesimals. Thanks to the continuous replenishment of the dust by
the planetesimal disk, the dust disk can be sustained over hundreds,
perhaps thousands, of Myr after the star has reached the MS. This
process slowly erodes the planetesimal disk and the dust content of
the disk is expected to decline with the star age as well. No clear
picture of the decline of the disk fractional luminosity $f_{\mathrm
d}=L_{\mathrm IR}/L_{\mathrm *}$ with time has nevertheless been
derived from the largest sample of debris disks studied so far, though
a threshold $f_{\mathrm d}$ value fading with stellar age can be
identified from the results of Decin et al.\,(2003). Observed dusty
disks may also be affected by the presence of planetary embryos. They
can stir up planetesimals through gravitational interaction thereby
increasing the rate of collisions and resulting in stochastic
brightness spikes of dust disks that may explain the observed large
spread of $f_{\mathrm d}$ as a function of stellar age (Kenyon \&
Bromley\,2002). Delayed stirring due to late planet formation has also
been proposed by Dominik \& Decin\,(2003) to explain this observed
spread in $f_{\mathrm d}$ at a given stellar age.

The observed $\mu$m-sized grains around MS stars are but the tip of
the iceberg of a size distribution that extends to planetesimal-sized
bodies in extra-solar planetary debris disks. Dusty disks are a
diagnostic for the presence of large solid bodies otherwise
undetectable, including planets that can under certain conditions
leave an imprint of their presence on the dust disk. Observed
structures in dusty disks could therefore serve to reveal and
characterize the mass and the orbital parameters of unseen planetary
companions. Several studies have recently explored this attractive
idea and have tried to constrain the distribution of the planetesimals
and the properties of possible perturber(s) from the observations of
the dust particles. I review in this paper current possible evidences
for a link between observed structures in dusty disks and the presence
of planets, starting with a brief summary of what we can learn from
the dusty Solar Sytem. The effect of stellar perturbers on
planetesimal gas-free disks is also discussed. I finally emphasize on
some theoretical and observational limitations.
\section{Structures in the Dusty Solar System}
\subsection{The Zodiacal Cloud}
IRAS satellite has revealed faint structures superimposed on the
broad-scale background zodiacal cloud. Among these structures, dust
bands inclined with respect to the ecliptic were evidenced by Low et
al.\,(1984) and confirmed later with the COBE/DIRBE and ISOPHOT
instruments in the mid-IR and in the visible from the ground. Low et
al.\,(1984) suggest that the bands are collisional debris within the
main asteroid belt between Mars and Jupiter rather than dust particles
released by short period comets while other structures like narrow
dust trails have a cometary origin. Some of these narrow trails don't
have identified parents and wide dust trails are still of unknown
origin (Sykes \& Walker\,1992).
The dust bands were thought to be associated with the three classical
Hirayama asteroid families but it has been recently proposed that the
Karin cluster and the Veritas family could be the only two sources of
dust (Nesvorn{\' y} et al.\,2003). Nesvorn{\' y} et al.\,(2003)
moreover propose that the dust bands result from recent (5-8\,Myr ago)
collisional disruptions of multi-km sized bodies.

The similarities between Jupiter's and the dust band particles'
inclinations and ascending nodes argue in favor of a dominant Jupiter
influence on their dynamics (Dermott et al.\,2001). The dusty Solar
System carries other signatures of the gravitational perturbation of
orbits of dust particles by the planets. The brightness enhancement of
the zodiacal cloud in the Earth trailing direction relative to the
leading direction can be explained by dust particles trapped into
resonant orbits while approaching the Earth (Dermott et
al.\,1994). This resonant trapping results in an asymmetric (in
azimuth) ring of particles with a cavity at the location of the Earth
and a trailing dust cloud. The trapped particles responsible for the
observed resonant ring are supposed to be mostly produced in the
asteroid belt. They slowly spiral towards the Sun because of the
Poynting-Robertson (PR) and the solar wind drag forces. The joint
effect of the drag forces and of the gravitational perturbation of the
planets could also explain other observed asymmetries in the zodiacal
cloud. In the framework of the secular perturbation theory (Murray \&
Dermott\,1999) and provided that the orbits of planets and grains are
not coplanar, the variation in semi-major axis of the orbit of the
dust particles directly translates in variation of the forced
inclinations of the orbit of the grains that depend both on the grain
size and on the heliocentric distance. The result is a warped zodiacal
dust disk (Wyatt et al.\,1999; Holmes\,2002), most likely due to
Jupiter
and Saturn
and which is indeed identified in the IRAS and COBE data sets (Deul \&
Wolstencroft\,1988, Dermott et al.\,1999).  Similarly, the forced
eccentricity of the dust particles due to Jupiter ($e=0.05$) provides
a theoretical frame to explain the observed offset of the center of
symmetry of the zodiacal cloud with respect to the Sun. But current
models do not readily account for this offset (Holmes et al.\,1998).
\subsection{A Kuiper Dust Disk?}
About $10^5$ objects larger than 100\,km are estimated to inhabit the
classical Kuiper Belt (KB) outside Neptune with semi-major axis $a$
between about 40 and 47\,AU and with a total mass of about
0.1\,M$_{\oplus}$ (Jewitt \& Luu\,2000). Whether the KB is in
collisional equilibrium and produces observable dust is currently not
known. According to Yamamoto \& Mukai\,(1998), low erosion of KB
objects (KBOs) by impacts of interstellar medium (ISM) dust particles
could contribute to the production of grains in a dusty counterpart of
the KB. An attempt to detect the thermal emission from the Kuiper dust
disk from COBE data failed because of the dominant zodiacal
contribution that could not be subtracted accurately enough
(Holmes\,2002). The existence of a Kuiper dust disk is therefore
neither theoretically nor observationally established, though the {\it
Pioneer} 10 and 11 spacecrafts sensitive to impacts caused by grains
larger than 10\,$\mu$m measured outside Saturn's orbit a flux of
grains coming from the KB at a rate consistent with the predictions of
Yamamoto \& Mukai\,(1998) (Landgraf et al.\,2002).

A significant fraction ($\sim$10\%) of the KBOs are trapped in
exterior mean motion resonances (MMRs) with Neptune. This includes the
Plutinos which are KBOs residing in the 3:2 MMR at $a\simeq
39.4$\,AU. The spatial distribution of resonant objects has a specific
azimuthally asymmetric structure that is a clear signature of the
ongoing dynamical process (e.g.\ Malhotra 1996). Dust particles
produced by collisional grinding of resonant objects or the trapping
of grains produced by non-resonant objects but migrating inward due to
PR drag could similarly harbor specific imprints that would indicate
the presence of Neptune. The theoretical shape and the observability
of a resonant Kuiper dust disk have been explored by Liou et
al.\,(1996, 1999), Moro-Mart{\'{\i}}n \& Malhotra\,(2002) and Holmes
et al. (2003) with the additional motive that identical structures
observed in extra-solar dusty disks could serve to reveal the presence
of exo-planets.

These studies show that only large grains stay long enough in the MMRs
with Neptune to have a chance to produce detectable signatures. Large
grains here means dust particles with $\beta$ ratios smaller than
about 0.1 (or grain radii larger than a few $\mu$m) where $\beta$ is
the ratio of radiation pressure to gravitation forces.  Holmes et
al.\,(2003) for instance considered the grains released by the
Plutinos and addressed the question of their ability to remain trapped
in the 3:2 MMR with Neptune by considering the action of PR drag,
radiation pressure, the solar wind drag, the planetary gravitational
perturbations but also the Lorentz force (since dust particles are
supposed to be positively charged) and the drag force due to the
neutral interstellar gas. They show that small grains produced by the
Plutinos with $\beta\geq 0.1$ have almost a zero probability to remain
trapped in the 3:2 MMR with Neptune. The observability of contrasted
structures in dusty disks produced by planetary perturbers therefore
relies on the size-frequency (or $\beta$-frequency) of dust
grains. For the KB, the size distribution is currently unknown while
that of the zodiacal cloud shows a peak in particle radius at about
50--100\,$\mu$m ($\beta<5\times10^{-3}$) at a distance of 1\,AU (Love
\& Brownlee\,1993).
\section{Observed Structures in Extra-Solar Dusty Systems}
\begin{figure}
\plotone{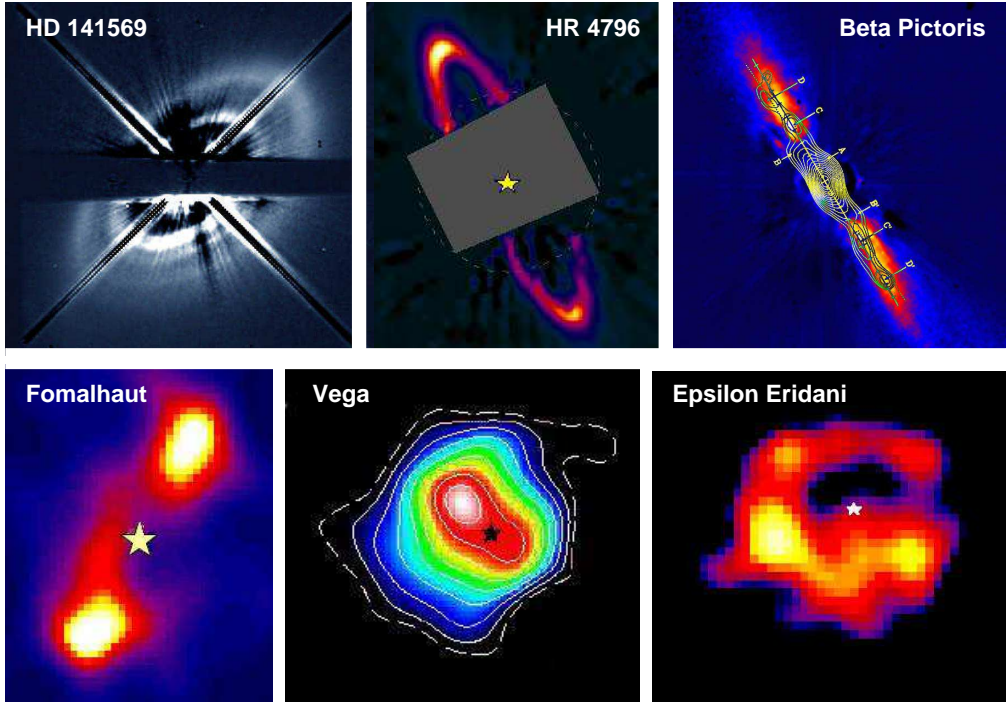}
\caption{Montage of resolved dusty disks around MS stars in ascending
  order from left to right, top to bottom (see also Tab.\,1).}
\end{figure}
Only a sparse set of dusty disks around MS stars has been spatially
resolved. Yet this sparse set should be considered as an impressive
improvement compared to the situation a decade ago when the 10$^{\sc
th}$ IAP Astrophysics meeting on ``Circumstellar Dust Disks and Planet
Formation'' was held in Paris (Proceedings edited by R. Ferlet \&
A. Vidal-Madjar). At that time $\beta$\,Pic was the only system for
which images were available. These images were revealing the
disk-shaped and flat geometry of a system seen almost edge-on and very
recent thermal images were suggestive of a clearing of the inner disk
unaccessible in scattered light (Pantin \& Lagage\,1994).

Fig.\,1 shows a montage of dusty disks firmly resolved so far around
MS stars in age ascending order and Tab.\,1 gives basic parameters for
the stars and for the disks discussed in this paper. The three top
panels of Fig.\,1 display disks seen in scattered light using
coronagraphic techniques to mask the central bright star. The visible
images of HD\,141569 and HR\,4796 have been obtained with the HST/STIS
(Mouillet et al.\,2001, Schneider et al.\,2001a) while the near-IR
image of $\beta$\,Pic has been obtained from the ground with adaptive
optics techniques (Mouillet et al.\,1997). The contours of the
deconvolved mid-IR image of the $\beta$\,Pic disk seen in thermal
emission at $\lambda$=$17.9\,\mu$m (Wahhaj et al.\,2003) are
superimposed on the near-IR image. The three oldest disks: Fomalhaut,
Vega and $\epsilon$\,Eri have only been resolved in thermal emission
with SCUBA in the sub-millimeter (bottom panels of Fig.\,1, Holland et
al.\,1998, 2003 and Greaves et al.\,1998) and with the Spitzer Space
Telescope (Stapelfeldt et al.\,2004, in prep.).
I refer to the paper by Lagage, these proceedings, for a complete
review on observations of dusty disks.
\begin{table}[t]
\caption{Resolved dusty disks from the youngest to the oldest. The
Table includes basic stellar parameters, a typical distance $R$ where
the disk shows a density peak and a typical radial width (FWHM) around
that distance, the observed inclination of the disk from pole-on, the
typical spatial resolution at which it has been best
resolved.
}
\begin{center}
\begin{tabular}{lccccccc} 
\tableline
Star & age & $d$ & Spectral  & R & $\Delta$R & $i$ & Res. \\ 
Name & [Myr] & [pc] & Type & [AU] & [AU] & [$^{\mathrm o}$] & [AU] \\
\tableline
HD\,141569 & 5$\pm$3 & 99 & B9.5V & {\begin{tabular}{c} 200 \\
310 \\
\end{tabular}}
& {\begin{tabular}{c} 45 \\ 120 \\
\end{tabular}} & 55 & 5 \\ 
HR\,4796 & 8$\pm$2 & 67 & A0V & 70 & 12 & 73 & 4 \\ 
$\beta$ Pic & 12$^{+8}_{-4}$ & 19.2 & A5V & 90 & 80 & 90 & 1 \\ 
Fomalhaut & 150$^{+200}_{-100}$ & 7.7 & A3V & 155 & 60 & 70 & 58 \\ 
Vega & 350$^{+30}_{-80}$ & 7.8 & A0V & 100 & 40 & 5 & 110 \\ 
$\epsilon$ Eri & 730$\pm$200 & 3.2 & K2V & 65 & 30 & 30 & 45 \\ 
\tableline
\tableline
\end{tabular}
\end{center}
\end{table}

The resolved dusty systems around MS stars display a wealth of
structures: ring-shaped disks accompanied with gaps (all the systems),
spiral structures and arcs (HD\,141569), clumps or blobs
($\beta$\,Pic, $\epsilon$\,Eri, Vega, Fomalhaut), offset asymmetries
(HD\,141569, HR\,4796), warps or offset inclinations ($\beta$\,Pic,
HR\,4796?, Fomalhaut?). Most of these structures can be seen in
Fig.\,1. Some of the systems are moreover expected to possess
exo-zodiacal dust populations, not observable on Fig.\,1, which are
deduced from spectral energy distribution modeling and/or mid-IR
resolved observations (HR\,4796, HD\,141569, $\beta$\,Pic,
Fomalhaut). An analogy with the structures described in Sec.\,1 and 2
is tempting. Similar structures in the dusty Solar System are indeed
related to the presence of the planets and the observed structures in
extra-solar dusty disks have raised the idea that they could as well
be due to yet undetected distant exoplanets in these systems
(Sec.\,4). It is interesting to note here that dusty disks have
currently only been resolved around early type stars except
$\epsilon$\,Eri.  Thus this approach is becoming an indirect but
complementary method to direct searches for exoplanets by radial
velocity and transit techniques that have mostly focused on nearby
solar-type stars and are sensitive to short-period planets. It should
also be noted that the two youngest systems (HR\,4796 and HD\,141569)
are both very likely members of multiple stellar systems in which case
planets may not be the only source of asymmetries in the dusty disks
(Sec.\,5).
%
\section{The Planetary Hypothesis}

%
%
Structures in dusty disks can be produced by resonant trapping of dust
grains with a gravitational perturber like a planet. Such an
explanation has been proposed for instance to explain the brightness
asymmetry at $\lambda=10$\,$\mu$m between the two extensions of the
edge-on $\beta$ Pic disk (Roques et al.\,1994). Several authors have
detailed the geometry of resonant signatures in a dusty disk
(e.g. recent papers by Ozernoy et al.\,2000, Kuchner \&
Holman\,2003). Lobes, arcs and voids are features that can serve to
locate and to characterize the type of dominant MMRs that depend in
particular on the mass of the gravitational perturber (Fig.\,2). These
approaches have provided possible explanations to some of the observed
structures in debris disks and observational tests have been proposed.

\begin{figure}
\plotone{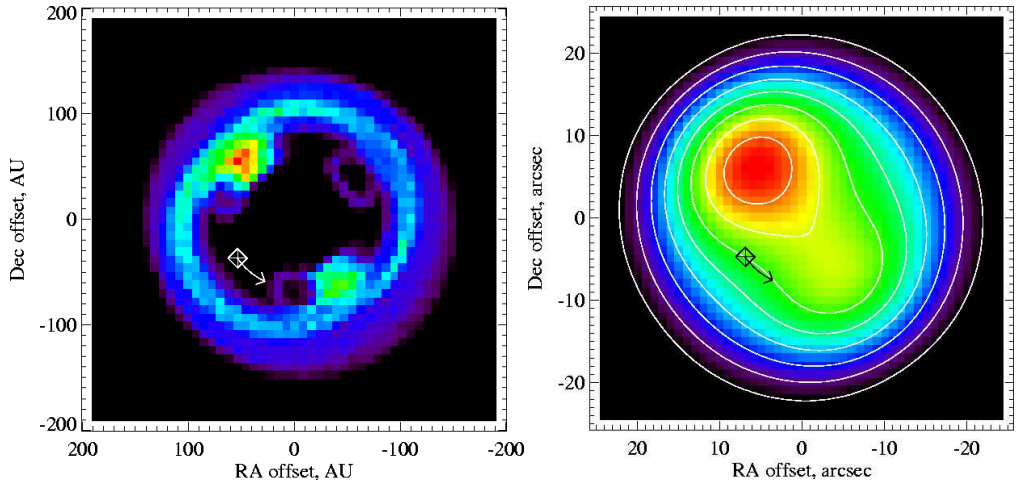}
\caption{Example of resonant structure in the Vega dusty disk (left)
and the corresponding synthetic sub-mm image (right)
(Wyatt\,2003). The square indicates the location of the planet (see
also Tab.\,2).}
\end{figure}

Resonant trapping with a planet can occur when the orbital period of
the planet is $(p+q)/p$ times that of a bound particle where $p$ and
$q$ are integers, $p>0$ and $p+q\geq1$. Kepler's third law implies:
$a/a_{\mathrm p}=[(p+q)/p]^{2/3}\times(1-\beta)^{1/3}$ where $a$ and
$a_{\mathrm p}$ are the semi-major axis of the orbit of the particle
and of the planet respectively and $\beta=F_{\mathrm rad}/F_{\mathrm
grav}$. Thus the location of an external MMR ($q>0$) moves inward with
$\beta$. Each resonance has a libration width $\Delta a$ around $a$
that depends on the eccentricity $e$ of the particle and on the planet
mass. In this libration zone, resonant orbits are stable
(e.g. Malhotra\,1996). But close to a planet the overlapping of
adjacent resonances results in a chaotic region.
\begin{table}[t]
\caption{Mass, semi-major axis and eccentricity of the predicted
planets in the $\epsilon$ Eri and Vega systems. The dominant MMRs are
indicated as well as the expected orientation of the planet on the sky
relative to the central star. The considered $\beta$ values are also
indicated. References: $^{(1)}$Ozernoy et al.(2000), $^{(2)}$Liou et
al.(2000), $^{(3)}$Quillen \& Thorndike(2002), $^{(4)}$Wilner et
al.(2002), $^{(5)}$Wyatt(2003).}
\begin{center}
\begin{tabular}{lccccccc}
\tableline
Star & M$_{\mathrm p}$ [M$_{\mathrm J}$]& $a_{\mathrm p}$ [AU] &
$e_{\mathrm p}$ & MMRs & $\beta$ & orientation & Ref. \\
\tableline
$\epsilon$ Eri & 0.2 & 55--65 & 0 & 2:1, 3:2 & 0.002 & West & $^{(1)}$ \\
$\epsilon$ Eri & 1 & 40 & 0.01 & 2:1, 3:2 & [0.002,0.05] & &
$^{(2)}$\\
$\epsilon$ Eri & 0.084 & 40 & 0.3 & 3:2, 5:3 & 0.1 & North & $^{(3)}$\\
\tableline
Vega & 2 & 50--60 & 0 & (1+$q$):1 & 0.3 & NW & $^{(1)}$ \\
Vega & 3 & 40 & 0.6 & (1+$q$):1 & 0.01 & NW & $^{(4)}$ \\
Vega & 0.054 & 60--70 & 0 & 2:1, 3:2 & 0 & NW or SE & $^{(5)}$ \\
\tableline
\tableline
\end{tabular}
\end{center}
\end{table}

Dust particles trapped in an outer resonance receive energy from the
inner planet and this energy can compensate the energy loss due to PR
drag. This predicts that external MMRs will last longer than internal
ones (Sicardy et al.\,1993, Murray \& Dermott\,1999) and current
studies of resonant structures in dusty disks have focused on MMRs
with $q>0$. This corresponds to a system with one planet and an outer
belt of planetesimals and dust that, in turn, resembles the
configuration of Neptune plus the KB in our Solar System
(Sec.\,2). Most models assume an unseen perturbing planet on a fixed
orbit, though Wyatt\,(2003) considered the outward migration of a
Neptune-like planet in the Vega system leading to an outer drift of
the resonant structure with time. Two types of dust trapping are
generally considered: particles produced in the MMRs by already
trapped parent bodies (e.g.\,Wyatt\,2003) or non-resonant particles
migrating inward due to PR drag and trapped in external MMRs
(e.g.\,Quillen \& Thorndike\,2002).

The results for $\epsilon$ Eri and Vega are summarized in Table 2.
Both systems require a planet with $a_{\mathrm p}\sim 40$--$70$\,AU
and the models agree on the rough location of the planet in the Vega
system.  But the different approaches used to derive the parameters of
the unseen planet, though some of them compare reasonably, do not
readily converge toward an unique solution. General trends can
nevertheless be drawn. The lumpy structure of the $\epsilon$ Eri ring
is generally better reproduced with a lower mass planet than for the
case of Vega. The dominant resonances are the first order 2:1 and 3:2
external MMRs but the 5:3 could also contribute significantly. Dust
trapping in these resonances results in arcs with asymmetric clumps
qualitatively matching the four-lobed structure of the $\epsilon$ Eri
ring. The two-lobed structure of the Vega ring can be reproduced with
a massive planet of a few Jupiter masses trapping dust particles in
the (1+$q$):1 external MMRs. Wyatt\,(2003) could nevertheless
reproduce the observations with a Neptune-like object (Fig.\,2). The
orbital periods of the planets being different from one model to
another, the resonant patterns should accordingly revolve at different
angular velocities. This provides an observational test to the
models. For instance Ozernoy et al.\,(2000) predict $\sim$0.7\deg/year
for the $\epsilon$ Eri structures whereas Quillen \& Thorndike\,(2002)
predict $\sim$1.3\deg/year.

Long-term planetary perturbations can also affect the structure of a
disk but, contrary to MMRs, they do not depend on the precise location
of the perturber on its orbit. The edge-on $\beta$\,Pic disk shows
vertical deformations that can primarily be explained by a planet on
an orbit inclined by a few degrees with respect to the initial
midplane of the disk. The precession of planetesimal orbits forces the
parent bodies of the observed dust grains to become coplanar with the
planet orbit. The precession frequency $\omega_{\mathrm p}$ decreases
with the distance from the central star. The vertical deformation
propagates outwards with time $t$ and stops approximately at the
distance for which $|\omega_{\mathrm p}^{-1}t|$ is $\sim$1. With an
assumed age of 20\,Myr, a 1\,M$_{\mathrm{Jup}}$ planet at 10\,AU does
produce a warp at $\sim$70\,AU as observed in the $\beta$\,Pic system
(Mouillet et al.\,1997).  The observable position of the warp shifts
only slightly outward when a size distribution of particles, their
dynamics and their optical properties are taken into account instead
of an unique population of planetesimals (Augereau et al.\,2001). Due
to radiation pressure the smallest bound grains with $\beta$ larger
than $\sim$0.15 are placed on orbits with significant $e$ values. They
spend a large fraction of their orbital time close to their apoastron
far from the ring of planetesimals where they originate from. These
small grains not only fill the outer regions of the disk and produce a
surface brightness consistent with the observations as expected, but
they also produce a vertical asymmetry at several hundreds of AU that
is a dusty counterpart of the planetesimal warp. In this model the
large-scale vertical (``butterfly'') asymmetry at hundreds of AU is
related to the presence of a planet at $\sim$10\,AU and requires a
colliding planetesimal ring peaked at about 90\,AU (Tab.\,1).
%
\section{Stellar Perturbers}
The three youngest systems shown in Fig.\,1 could have been affected
by stellar companions. The sharply defined ring around HR\,4796 for
instance shows a faint brightness asymmetry in thermal emission (at
$\lambda\sim20\,\mu$m) and in scattered light which could be due to
the perturbation of the disk by a close M companion thought to be
bound. Provided that this stellar companion, located at a projected
distance of 517\,AU from HR\,4796, is on an eccentric orbit ($e\sim
0.13$), the dust particles suffer a small but sufficient forced
eccentricity. The result is a ring of dust offset from the central
star and a brightness enhancement of the dust near the forced
pericenter of the perturbed disk (Wyatt et al.\,1999).

The secular perturbation of the HD\,141569 disk by the two stellar M
companions also provides an explanation for some of the observed
asymmetries and for the size of the disk. Assuming that at least one
of the companions is bound with HD\,141569 and on a sufficiently
eccentric orbit, it excites a spiral density wave which qualitatively
matches the observed asymmetric ring of dust particles in the outer
regions of the disk (at $\sim$325\,AU) after the perturbers have only
completed $\sim$10 orbital revolutions (Fig.\,3, Augereau \&
Papaloizou,\,2004a). This corresponds to a disk evolution timescale of
a few Myr if $e$ is between 0.7 and 0.9 with the derived pericenter
distance of 930\,AU. In that approach the wide dark lane between the
two asymmetric rings at 200\,AU and 325\,AU is not regarded as a
depleted region which would suggest a mechanism to clean up the dust
disk in that region. Rather the two asymmetric rings are considered as
two independent coherent over-densities produced by perturbers: the
stellar companions for the outermost structure and probably substellar
object(s) for the inner ring.

\begin{figure}
\plotone{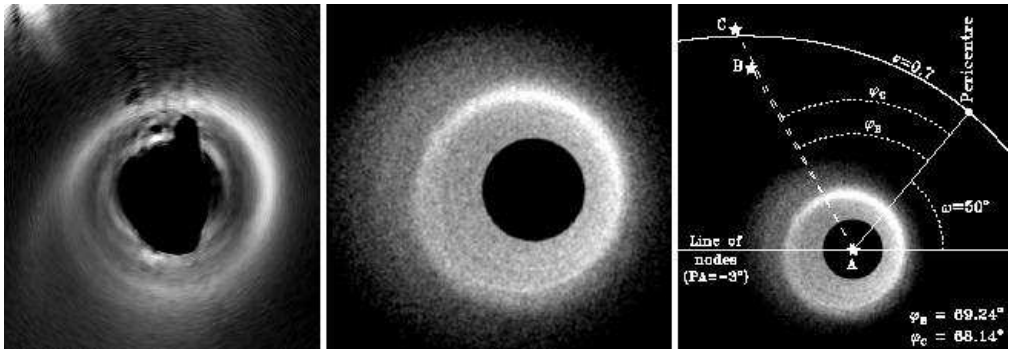}
\caption{Middle and right panels: model proposed by Augereau \&
Papaloizou\,(2004a) to reproduce the observed spiral structure in the
outer regions of the HD\,141569 disk (left panel, Clampin et
al.\,2003).}
\end{figure}
Passing stars could also leave an imprint on a disk.  According to
Kalas et al.\,(2000), the $\beta$\,Pic disk could have recently
experienced a low relative velocity encounter with a 0.5\,M$_{\odot}$
star on a low-inclination and parabolic orbit. Such a stellar flyby
produces near to the periastron of the stellar perturber, at about
700\,AU from the central star, a transient spiral structure that
collapses in eccentric ring-like density concentrations in
$\sim$0.1\,Myr. The concentrations, when seen edge-on, coincide with
the positions of faint features that appear in one of the two
extensions of the disk at distances between 500 and 785\,AU from the
star when a smooth scattered light disk model is subtracted from the
observations.  The length asymmetry of the disk can also be
reproduced. If the flyby is not coplanar, the model proposed by Kalas
et al.\,(2000) provides an alternative explanation to the vertical
(warp and butterfly) asymmetries in the disk (see also Sec.\,4). No
convincing candidate star in the neighborhood of $\beta$\,Pic has been
identified so far which presently makes this approach unlikely.
%
%
\section{Current Modeling Limitations and Alternative Approaches}
Several models currently proposed to reproduce the observed structures
in dusty disks rely on the trapping of particles in MMRs with a planet
in collision-less systems or in systems where the collision timescale
of particles is larger than their resident timescale in a MMR. Too
frequent collisions could dissolve structures. Lecavelier et
al.\,(1996) have shown that resonant structures in the $\beta$\,Pic
disk are observable at a few tens of AU if the surface density remains
below a critical value that translates into a maximal vertical optical
thickness of a few $10^{-4}$. Holmes et al.\,(2003), addressing the
question of the observability of a Plutino dust disk, clearly show
that contrasted structures only appear for large particles (small
$\beta$ values) since their dispersion in semi-major axis after a
collision remains generally smaller than the libration amplitude of
the resonance. As a consequence images dominated by particles with
large $\beta$ values should hardly show resonant structures.

The situation may be worsened by the effect of the radiation pressure.
The trapping of a dust particle in an external MMR results in a
raising of its eccentricity on a characteristic timescale that drops
with $\beta$ (Liou \& Zook\,1997). Thus $e$ raises faster with
$\beta$. In other words, radiation pressure can speed up the ejection
of a particle from a MMR if $e$ can reach the critical maximum value
of the eccentricity allowed in that resonance (Weidenschilling \&
Jackson\,1993).  The sharpness of resonant features therefore
critically depends on $\beta$ and, in turn, on the grain size
distribution in the disk. Recent studies have shown that the size
distribution resulting from collisions in a disk with a lower cutoff
size (due for instance to radiation pressure) could depart
significantly from the theoretical collisional equilibrium
distribution described by the classical -3.5 power law
(e.g. Th{\'e}bault et al.\,2003 and ref.\ therein). The size-strength
law of solid bodies for describing collisions is moreover badly
constrained. Finally, the contribution of evaporating comet-like
bodies to dust disks is also not known.

Several processes not discussed here could also affect structures in
dusty disks. As indicated in Sec.\,2.2, impacts from ISM grains could
contribute to the production of dust in the Solar System and may have
the ability to erase any signature of a planet in the outer Solar
System (Moro-Mart{\'{\i}}n \& Malhotra\,2002). This effect could
nevertheless be negligible for earlier-type stars (A-F) because of the
relatively strong radiation pressure (Artymowicz \&
Clampin,\,1997). Stellar wind and stellar magnetic field can also be
considered. Interplanetary dust particles for instance are charged by
the emission of photoelectrons due to solar UV and to solar wind ions
(Kempf et al.\,2001). The Lorentz force may not be negligible for the
smallest and closest particles (e.g. Barge et al.\,1982, Fahr et
al.\,1995). The temporal and spatial variations of $\vec B$ in the
Solar System have been shown to affect the orbits of charged dust
particles by causing a random walk in semi-major axis and a dispersion
in inclination and a precession of nodes. Stochastic collisions are
another potential source of asymmetries in disks. It has been shown
nevertheless that the clumps in the Fomalhaut disk are unlikely due to
that mechanism (Wyatt \& Dent\,2002). Alternatively, gas-dust coupling
can structure the youngest disks when the gas is not yet entirely
dissipated (Takeuchi \& Artymowicz, 2001). Gas in Keplerian rotation
has indeed been observed in the HD\,141569 and $\beta$\,Pic systems
(Augereau et al.\,2004b, Brandeker et al.\,2004)
\section{Future Prospects}
Inferring the presence of planets from structures in dusty disks
presently suffers from a major problem: the set of spatially resolved
dusty disks is still sparse. But it is noteworthy that well marked
asymmetries and annular shapes have been systematically
observed. Therefore detecting unseen planetary companions from
perturbed disks geometries is a promising approach while it remains in
its infancy mostly because of observational limitations.

The three oldest systems for instance (Fig.\,1) have only been
resolved at very low spatial resolution, blurring possible fine
structures produced by embedded planets. Two different millimeter
interferometers, sensitive to higher spatial frequency structures than
the sub-mm SCUBA images, could resolve two lobes in the Vega disk but
curiously the lobes do not properly match each other (Koerner et
al.\,2001, Wilner et al.\,2002). Future instruments like CARMA and
ALMA for instance will have access to optically thin disks that are
borderline targets for current millimeter interferometers and will be
able to provide constraints on the residual gas. The direct search for
planets around stars with debris disks has to this date not yet
resulted in any detection and no such debris disk has been resolved so
far around those stars for which giant planets were detected by radial
velocity techniques.  The $\epsilon$\,Eri system (Fig.\,1) is an
exception since it could host a Jupiter-mass planet (Hatzes et
al.\,2000). Presumed detections of disks around stars with known
planets have been claimed but almost all of them are awaiting a
confirmation or have been invalidated by follow-up observations
(e.g. 55\,Cnc, Schneider et al.\,2001b).

The handful of resolved disks also contrasts with the number of
expected debris disks in the solar neighborhood: 20\% on average and
up to 40\% for A type stars (Habing et al.\,2001). These statistics
and the time-dependant evolution of the disk luminosity will be
revisited soon with the {\sc Spitzer} Space Telescope. These
observations will provide a valuable database for the search of faint
extended and structured emissions around MS stars with future high
resolution and high contrast instruments on ground-based telescopes
(e.g.\ VLT/Planet-Finder) and on-board the JWST. By constraining the
dust content and the structure of the dusty disks, these observations
will also help to prepare future missions such as TPF/Darwin that aim
at directly detecting Earth-like planets.
\acknowledgments I am very grateful to V.C. Geers for carefully
reading the manuscript. This work is supported by the European
Community's Human Potential Programme under contract
HPRN-CT-2002-00308, PLANETS.

\end{document}